\documentclass[aip,amsmath,amssymb,reprint]{revtex4-1}
\usepackage{graphicx}
\usepackage{dcolumn}
\usepackage{bm}

\usepackage[utf8]{inputenc}
\usepackage[T1]{fontenc}
\usepackage{mathptmx}
\usepackage{epstopdf,mathrsfs}
\usepackage[colorlinks,linktocpage,linkcolor=blue]{hyperref}

\draft
\begin{document}

\preprint{AIP/123-QED}

\title{Ergodic properties of heterogeneous diffusion processes in a potential well}

\author{Xudong Wang}
 \altaffiliation{}
\author{Weihua Deng}
\email{dengwh@lzu.edu.cn}
\altaffiliation{}
\author{Yao Chen}
\affiliation{%
School of Mathematics and Statistics, Gansu Key Laboratory
of Applied Mathematics and Complex Systems, Lanzhou University, Lanzhou 730000,
P.R. China
}%


\begin{abstract}
Heterogeneous diffusion processes can be well described by an overdamped Langevin equation with space-dependent diffusivity $D(x)$. We investigate the ergodic and non-ergodic behavior of these processes in an arbitrary potential well $U(x)$ in terms of the observable---occupation time. Since our main concern is the large-$x$ behavior for long times, the diffusivity and potential are, respectively, assumed as the power-law forms $D(x)=D_0|x|^\alpha$ and $U(x)=U_0|x|^\beta$ for simplicity. Based on the competition roles played by $D(x)$ and $U(x)$, three different cases, $\beta>\alpha$, $\beta=\alpha$, and $\beta<\alpha$, are discussed. The system is ergodic for the first case $\beta>\alpha$, where the time average agrees with the ensemble average, being both determined by the steady solution for long times. In contrast, the system is non-ergodic for $\beta<\alpha$, where the relation between time average and ensemble average is uncovered by infinite-ergodic theory. For the middle case $\beta=\alpha$, the ergodic property, depending on the prefactors $D_0$ and $U_0$, becomes more delicate. The probability density distribution of the time averaged occupation time for three different cases are also evaluated from Monte Carlo simulations.
\end{abstract}

\maketitle

\section{Introduction}
The anomalous diffusion phenomena are ubiquitous in the natural world, especially in numerous microscopic systems. Over the last two decades, much effort has been devoted to the study of anomalous diffusion phenomena \cite{HausKehr:1987,BouchaudGeorges:1990,MetzlerKlafter:2000}, characterized by nonlinear time dependence of mean squared displacement (MSD)
\begin{equation} \label{eq1}
  \langle x^2(t)\rangle \simeq 2K_\gamma t^\gamma,
\end{equation}
where $\gamma\neq1$ is the anomalous diffusion exponent and $K_\gamma$ is the anomalous diffusion coefficient with physical dimension $[\textrm{cm}^2/\textrm{s}^\gamma]$; it is called subdiffusion for $0<\gamma<1$ and superdiffusion for $\gamma>1$.

However, the MSD is not the only criterion to distinguish the processes, i.e., some fundamentally different processes may have the same MSD. For example, the subdiffusion was first observed in amorphous semiconductors described by continuous-time random walk (CTRW) models \cite{ScherMontroll:1975}, being Non-Gaussian process, while it was also observed in the class of viscoelastic system described by the generalized Langevin equation with (tempered) power-law friction kernel \cite{Lutz:2001,Goychuk:2012,SlezakMetzlerMagdziarz:2018,GarciaSandevSafdariPagniniChechkinMetzler:2018,DengBarkai:2009} and of (tempered) fractional Brownian motion \cite{MandelbrotNess:1968,MeerschaertSabzikar:2013,ChenWangDeng:2017}, both of which are Gaussian processes. There are also many different superdiffusion processes, such as, L\'{e}vy flight \cite{ShlesingerZaslavskyFrisch:1995}, L\'{e}vy walk \cite{ZaburdaevDenisovKlafter:2015}, and those in the underdamped Langevin systems \cite{EuleFriedrichJenkoKleinhans:2007,WangChenDeng:2018-2}.
In recent years, more and more dynamics in heterogeneous systems can be well described by the overdamped Langevin equation with space-dependent diffusivity \cite{CherstvyChechkinMetzler:2013,CherstvyMetzler:2013,CherstvyMetzler:2014}
\begin{equation}\label{Int1}
  \frac{dx}{dt}=\sqrt{2D(x)}\xi(t),
\end{equation}
such as the Richardson diffusion in turbulence \cite{Richardson:1926}, mesoscopic approaches to transport in heterogenous porous media \cite{HaggertyGorelick:1995,DentzGouzeRussianDweikDelay:2012} and on random fractals \cite{Loverdo.etal:2009,OShaughnessyProcaccia:1985}.
The heterogeneous dynamical behavior is particularly remarkable in biological systems.
The cytoplasm of biological cells is always crowded with various obstacles, including proteins, nucleic acids, ribosomes, the cytoskeleton, as well as internal membranes compartmentalizing the cell \cite{ZimmermanMinton:1993,ZhouRivasMinton:2008}. The nonuniform distribution of crowders in the cytoplasm provides the heterogenous media for tracer particles of different sizes in it.
The motion of the tracer particles might be imposed by the cage effects if the size of the particle is comparable to the local mesh size in the crowded environments \cite{Wong.etal:2004,GodecBauerMetzler:2014,LeeCrosbyEmrickHayward:2014}.
This kind of heterogeneous diffusion process also can be realized in the experiments on eukaryotic cells \cite{Kuhn.etal:2011} and a local variation of the temperature in thermophoresis experiments \cite{MastSchinkGerlandBraun:2013,MaedaTlustyLibchaber:2012}.

Nowadays, single particle tracking has become a powerful tool to study transport processes in cellular membranes \cite{SaxtonJacobson:1997} and probe the microrheology of the cytoplasm \cite{Wirtz:2009,YaoTassieriPadgettCooper:2009}. It can be used to evaluate the time averaged observables in cells through video microscopy of fluorescently labeled molecules. The time average is equal to the ensemble average for an ergodic system. But for anomalous diffusion, especially for the molecules diffusing in living cells, the time average often becomes a random variable and irreproducible for the individual trajectories. For the heterogeneous diffusion process described by Langevin equation \eqref{Int1} with the diffusivity $D(x)$ in the power-law form \cite{Gardiner:1983,Srokowski:2007}:
\begin{equation}\label{diffusivity}
  D(x)=D_0|x|^\alpha, \qquad \alpha \leq 2,
\end{equation}
it is subdiffusion for $\alpha<0$ and superdiffusion for $\alpha>0$; the weak non-ergodic behavior has also been observed in \cite{CherstvyChechkinMetzler:2013,CherstvyMetzler:2013,CherstvyMetzler:2014}, where the time averaged MSD remains linear in the lag time and thus differs from the corresponding ensemble average.

Here we consider ergodic/non-ergodic properties of the heterogeneous diffusion processes \eqref{Int1} in a potential well
\begin{equation}\label{well}
  U(x)=U_0|x|^\beta.
\end{equation}
The non-ergodic behavior of subdiffusive CTRW model in a confined potential has been investigated in \cite{BelBarkai:2006}.
Even for the ergodic system where normal (anomalous) diffusion modeled by (fractional) Brownian motion and the (fractional) Langevin equation is confined by a harmonic potential, the time averaged MSDs behave differently from the ensemble averaged counterparts \cite{JeonMetzler:2012}. We here investigate the ergodic behavior in terms of the occupation time $T_{\mathcal{D}}(t)$ for a particle staying in a certain domain $\mathcal{D}$. It is found that the physical mechanism of the heterogeneous diffusion process in \eqref{Int1} is different from CTRW model and fractional Langevin equation, although they may have the same anomalous diffusion exponents defined in (\ref{eq1}). If the potential well \eqref{well} is deep enough, the system might reach a steady state and the ergodic hypothesis is shown to be valid.

However, for a potential well that is not very deep, the steady solution is found to be non-normalizable, implying that the non-trival steady solution may not exist. Non-normalizable steady solutions have been previously discussed in the literature on physical issues, e.g., L\'{e}vy walk and laser-cooled atoms \cite{KesslerBarkai:2010,LutzRenzoni:2013,RebenshtokDenisovHanggiBarkai:2014,AghionKesslerBarkai:2017}, the weakly-chaotic maps \cite{KorabelBarkai:2009,AkimotoMiyaguchi:2010}, nonlinear oscillators \cite{MeyerKantz:2017}, the thermodynamics \cite{AghionKesslerBarkai:2018}, and multiplicative noise \cite{LeibovichBarkai:2018}.
From these discussions, one can note that the (non-normalizable) steady solution is determined by the competition role played by the potential $U(x)$ and the space-dependent diffusivity $D(x)$ in our concerned model. A generalized form of ergodicity, namely infinite-ergodic theory \cite{Aaronson:1997,AaronsonThalerZweimuller:2005,ThalerZweimuller:2006}, applies to the non-normalizable case, which uncovers the relation between time averages and ensemble averages and helps to characterize the distribution of time averages.

This paper is organized as follows. In Sec. \ref{Sec2}, we briefly review the heterogeneous diffusion processes, especially their probability density functions (PDFs), and then turn to their steady solutions under an arbitrary potential well. In Sec. \ref{Sec3}, the system with a deep potential (i.e., $\beta>\alpha$) is explicitly discussed; in this case, the steady solution is normalizable, and the distribution of time averaged occupation time is a $\delta$-function centered on the value of the ensemble average, which is determined by the steady solution for long times.
In the case of comparable potential well (i.e., $\beta=\alpha$) and weak potential well (i.e., $\beta<\alpha$), where the steady solution is non-normalizable, the infinite-ergodic theory and the distribution of the time averages are discussed in Sec. \ref{Sec4} and Sec. \ref{Sec5},  respectively. A summary of the key results is provided in Sec. \ref{Sec6}. In the appendices some mathematical details are collected.

\section{Heterogeneous diffusion model}\label{Sec2}
\subsection{No potential well}

We firstly review the stochastic Langevin model in a heterogeneous medium with space-dependent diffusivity $D(x)$, namely,
\begin{equation}\label{model1}
  \frac{dx}{dt}=\sqrt{2D(x)}\xi(t),
\end{equation}
where $\xi(t)$ is a Gaussian white noise with zero mean $\langle \xi(t)\rangle$=0 and $\delta$-correlation $\langle \xi(t)\xi(t')\rangle=\delta(t-t')$. Here the multiplicative noise term $\sqrt{2D(x)}$ is interpreted in the Stratonovich sense \cite{Risken:1989}. For a general stochastic differential equation, the Stratonovich prescription has the advantage of ordinary chain rule formulas under a transformation \cite{Risken:1989,Oksendal:2005} and of yielding physically correct results especially for the noise with infinitely short correlation times \cite{WestBulsaraLindenbergSeshadriShuler:1979}, while the It\^{o} prescription is mathematically and technically the most satisfactory  \cite{Gardiner:1983} and easy for simulations.

During the analytical calculations, the specific form of $D(x)$ in \eqref{diffusivity} is taken.
It is actually regularized at $x=0$ in physical systems and in simulations \cite{CherstvyChechkinMetzler:2013}. For $\alpha>0$, a modified form
  $D(x)=D_0(A_0+|x|^\alpha)$ with a positive constant $A_0$ is
to prevent the particles trapping at $x=0$. For $\alpha<0$, we take
  $D(x)=\frac{D_0A_0}{A_0+|x|^{-\alpha}}$
to avoid the divergences at $x=0$. On the other hand, the regularization at $x=0$ makes it satisfy  the Lipschitz condition and the request $\alpha\leq2$ ensures the growth condition for existence and uniqueness of the solution of the Markovian stochastic differential equation \eqref{model1} \cite{Oksendal:2005}.
By the substitution \cite{CherstvyChechkinMetzler:2013}
\begin{equation}\label{substitution}
\begin{split}
    y&=\int^x \frac{1}{\sqrt{2D(x')}} dx'   \\
    &= \frac{\sqrt{2/D_0}}{2-\alpha}|x|^{\frac{2-\alpha}{2}}\,\textrm{sign}(x),
\end{split}
\end{equation}
where $y(t)$ is the standard Brownian motion \cite{Risken:1989}, the PDF and ensemble averaged MSD of $x(t)$ for $\alpha<2$ are obtained in \cite{CherstvyChechkinMetzler:2013}
\begin{equation}\label{pdf1}
  p(x,t)=\frac{|x|^{-\alpha/2}}{\sqrt{4\pi D_0t}} \exp\left(-\frac{|x|^{2-\alpha}}{(2-\alpha)^2D_0t}\right)
\end{equation}
and
\begin{equation}\label{less2}
  \langle x^2(t)\rangle \propto t^{2/(2-\alpha)},
\end{equation}
which is obviously superdiffusion for $\alpha>0$ and subdiffusion for $\alpha<0$.
For the critical value $\alpha=2$, by the same substitution \eqref{substitution}, one can obtain the relation between $y(t)$ and $x(t)$:
\begin{equation}
  y(t)=(2D_0)^{-1/2}\textrm{sign}(x) \ln|x|.
\end{equation}
Therefore, the PDF and ensemble averaged MSD of $x(t)$ for $\alpha=2$ are, respectively,
\begin{equation}\label{pdf2}
  p(x,t)=\frac{1}{\sqrt{4\pi D_0 t}} |x|^{-1-\frac{\ln|x|}{4 D_0t}}
\end{equation}
and
\begin{equation}
  \langle x^2(t)\rangle \propto e^{4 D_0t},
\end{equation}
which grows exponentially and much faster than \eqref{less2}.
The case of $\alpha=2$ is common in applications; more often it emerges together with a harmonic potential. In this case, the Langevin system can be interpreted as the model for a massless particle connected to a spring with the spring coefficient varying randomly in time \cite{Deutsch:1994}, which also is used to characterize the L\'{e}vy-type stochastic processes in terms of continuous trajectories of walker motion \cite{LubashevskyFriedrichHeuer:2009_1}. Note that the PDFs \eqref{pdf1} and  \eqref{pdf2} decay faster than any power-law form, implying that the condition $\alpha\leq2$ indeed guarantees the existence of the solution of Langevin equation \eqref{model1}.

\subsection{Steady solution in a potential well}
We assume that the potential well $U(x)=U_0|x|^\beta$ with $U_0>0$, in which the exponent $\beta$ is usually positive; especially $\beta=2$ is for the potential between the monomers in a bead spring polymer chain
\cite{ShinCherstvyKimMetzler:2015}. Sometimes, the exponent $\beta$ might also be negative (e.g., the Lennard-Jones potential \cite{ShinCherstvyKimMetzler:2015,ShinCherstvyMetzler:2015} approximating the interaction between a pair of neutral atoms or molecules) and in this case we regularize $U(x)$ at the origin by the way similar to $D(x)$. Then the force is $f(x)=-dU(x)/dx=-U_0\beta|x|^{\beta-1}\textrm{sign}(x)$ and the Langevin equation for a particle in this potential well is presented as
\begin{equation}\label{model2}
  \frac{dx}{dt}= f(x) +\sqrt{2D(x)} \xi(t),
\end{equation}
the Fokker-Planck equation \cite{Risken:1989,CherstvyChechkinMetzler:2013} of which for the PDF $p(x,t)$ is
\begin{equation}\label{FFP}
    \frac{\partial p(x,t)}{\partial t}= \mathcal{L}_{\textrm{FP}} ~p(x,t)
\end{equation}
with the Fokker-Planck operator
\begin{equation}
  \mathcal{L}_{\textrm{FP}} =-\frac{\partial}{\partial x}f(x) + \frac{\partial}{\partial x}\sqrt{D(x)}\frac{\partial}{\partial x}\sqrt{D(x)}.
\end{equation}
This equation can be represented through probability current $J(x,t)$ due to mass conservation \cite{Risken:1989} as $\partial_tp(x,t)=-\partial_xJ(x,t)$ with
\begin{equation}
  J(x,t)=f(x)p(x,t)-\sqrt{D(x)}\frac{\partial}{\partial x}\sqrt{D(x)}p(x,t).
\end{equation}
In many circumstances, the steady solution $p^{st}(x)$ of \eqref{model2} can be directly obtained by setting $J(x,t)=0$, which yields
\begin{equation}\label{pst0}
  p^{st}(x)= \frac{N}{\sqrt{D(x)}}\exp\left(\int^x\frac{f(x')}{D(x')}dx'\right),
\end{equation}
with $N$ being the normalization constant.
It can be seen that the strengths of $f(x)$ and $D(x)$ dominate the shape of $p^{st}(x)$ in \eqref{pst0}. For different exponents of the power-law form of the potential $U(x)$ and diffusivity $D(x)$, the solution $p^{st}(x)$ behaves as
\begin{equation}\label{pst}
  \begin{split}
    p^{st}(x)=
    \begin{cases}
      \frac{N}{\sqrt{D_0}}|x|^{-\alpha/2}\exp\left(-\frac{U_0\beta}{D_0(\beta-\alpha)}|x|^{\beta-\alpha}\right) & \beta\neq\alpha \\[5pt]
      \frac{N}{\sqrt{D_0}}|x|^{-\alpha/2-U_0\beta/D_0}                                                          & \beta=\alpha .
    \end{cases}
  \end{split}
\end{equation}
Here it should be emphasized that the real diffusivity $D(x)$ in \eqref{diffusivity} is regular at origin. So the PDF $p^{st}(x)$ should also be regular at origin. What we pay attention to in this paper is the large-$x$ behavior of the solution $p^{st}(x)$ in \eqref{pst}.

For the case of $\beta>\alpha$ and $\beta>0$, $p^{st}(x)$ decays exponentially for large $x$ and it can be normalized whatever other parameters are. After some calculations, the normalization constant $N$ can be obtained as
\begin{equation*}
  N=\frac{\sqrt{D_0}(\beta-\alpha)}{2\Gamma(\varrho)} \left(\frac{U_0\beta}{D_0(\beta-\alpha)}\right)^\varrho
\end{equation*}
with $\varrho=(2-\alpha)/(2(\beta-\alpha))$. For the other two cases $\beta=\alpha$ and $\beta<\alpha$, however, when $\alpha>0$, the PDF $p^{st}(x)$ decays as power-law 
 $|x|^{-\alpha/2-U_0\beta/D_0}$ and $|x|^{-\alpha/2}$, respectively,
which even can not be  normalized for some parameters. The divergence at infinity means that the steady state may not exist in the usual sense and a time-dependent solution should be considered to measure some observables.

In the following, we will use the steady solution \eqref{pst} to investigate the time and ensemble averages of some observables as well as the ergodic properties of the Langevin system \eqref{model2}.
It has been shown that the steady solution is completely different for different magnitudes of $\beta$ and $\alpha$. For the first case of $\beta>\alpha$ and $\beta>0$, the steady solution can be normalized even for $\alpha<0$. For the other two cases $\beta=\alpha$ and $\beta<\alpha$, the condition $\alpha>0$ is assumed; otherwise, the steady solution grows with time for large $x$, which is impossible for a PDF.
We mainly explore the occupation time statistic $T_{\mathcal{D}}(t)$ in a given domain $\mathcal{D}$ for three different cases,  respectively, and then make some comparisons. The concept of occupation time has been studied by probabilists for a long time \cite{Lamperti:1958}; in recent years, it attracts interests of physicists together with some related observables, such as first-passage time \cite{Redner:2001}, persistence \cite{MajumdarRossoZoia:2010}, and the area under Brownian and non-Brownian path \cite{Grebenkov:2007,MajumdarComtet:2005,BarkaiAghionKessler:2014}.

\section{Ergodic phase for $\beta>\alpha$}\label{Sec3}
In the case of $\beta>\alpha$, we claim that for large $t$ the PDF of the occupation fraction
\begin{equation}
  \bar{p}_{\mathcal{D}}= \frac{T_{\mathcal{D}}(t)}{t}
\end{equation}
satisfying
\begin{equation}\label{pdf3}
  f(\bar{p}_{\mathcal{D}})=\delta(\bar{p}_{\mathcal{D}}-P^{st}_{\mathcal{D}}),
\end{equation}
where $P^{st}_{\mathcal{D}}=\int_{\mathcal{D}} p^{st}(x)dx$ is the ensemble average in the steady state;
this kind of phenomenon has been ever detected in \cite{BelBarkai:2006}, where the PDF of $\bar{p}_{\mathcal{D}}$ for a certain CTRW model is
\begin{equation}\label{pdf4}
\begin{split}
    &f_{\textrm{CT}}(\bar{p}_{\mathcal{D}}) =\frac{\sin(\mu\pi)}{\pi} \times \\
    &~~
  \frac{\mathcal{R}_{\mathcal{D}} \bar{p}_{\mathcal{D}}^{\mu-1}(1-\bar{p}_{\mathcal{D}})^{\mu-1}}
                                    {\mathcal{R}_{\mathcal{D}}^2(1-\bar{p}_{\mathcal{D}})^{2\mu}+\bar{p}_{\mathcal{D}}^{2\mu}+2\mathcal{R}_{\mathcal{D}}(1-\bar{p}_{\mathcal{D}})^{\mu}\bar{p}_{\mathcal{D}}^{\mu}\cos(\mu\pi)}
\end{split}
\end{equation}
with $\mu$ being the exponent of the power-law distributed waiting times and
\begin{equation}
  \mathcal{R}_{\mathcal{D}}=\frac{P^{st}_{\mathcal{D}}}{1-P^{st}_{\mathcal{D}}}.
\end{equation}
When $\mu=1$, the PDF of $\bar{p}_{\mathcal{D}}$ in \eqref{pdf4} reduces to the ergodic phase \eqref{pdf3}.

Now we want to derive the PDF $f(\bar{p}_{\mathcal{D}})$ in \eqref{pdf3} for our model (in Langevin picture) based on the CTRW framework in \cite{BelBarkai:2006}. It was proposed by Fogedby \cite{Fogedby:1994}  that an overdamped Langevin equation (with additive Gaussian noise) coupled with a subordinator can describe the process defined by  CTRW model, being the same one after taking scaling limit.
 But it seems not very clear what the multiplicative noise in our model \eqref{model2} underlies in a CTRW model.
Some papers \cite{FedotovFalconer:2012,Srokowski:2009,KaminskaSrokowski:2018} derived the generalized Fokker-Planck equation from CTRW model with variable jumping rate, which shows that the space-dependent diffusivity may come from space-dependent distribution of waiting time.
Here, we present an alternative method to clarify a clear relation between the space-dependent diffusivity in a Langevin equation and the distribution of waiting times in CTRW framework (see the detailed derivation in Appendix \ref{APPa}). The multiplicative noise term $\sqrt{2D(x)}$ in \eqref{model2} implies the exponential distributed waiting time with rate $2D(x)$ in CTRW model. The bigger $D(x)$ is, the larger number of renewals per unit time gets and the faster the diffusion becomes.
Note that even the exponent of the power-law form of $D(x)$ decides the diffusion behavior of the particle in \eqref{model1} (superdiffusion for $\alpha>0$ and subdiffusion for $\alpha<0$), the mechanism is, respectively, completely different from the superdiffusive L\'{e}vy flight \cite{ShlesingerZaslavskyFrisch:1995} and subdiffusive CTRW with power-law distributed waiting times \cite{HausKehr:1987,BouchaudGeorges:1990,MetzlerKlafter:2000}. 
For the system \eqref{model1}, the second moment of the jump lengths and the mean of waiting times are all finite, moreover the latter is space-dependent; this interpretation from CTRW viewpoint  plays an important role in deriving and understanding the result \eqref{pdf3}, which will be shown immediately.

The distribution of the occupation time of a Brownian motion or non-Brownian motion can be derived by probability theory \cite{Levy:1939} or by backward Feynmann-Kac approach \cite{Kac:1949,Majumdar:2005,TurgemanCarmiBarkai:2009,WangChenDeng:2018}.
The form of the PDF $f(\bar{p}_{\mathcal{D}})$ in \eqref{pdf4} has been derived by Lamperti \cite{Lamperti:1958} based on the mathematical theory of occupation times and rederived by Bel and Barkai \cite{BelBarkai:2006} in the statistical physics community for a two-state process.  Here, we also consider a two-state process with the two states:  
the occupation times $T_+$ and $T_-$ in the positive and negative half-space, respectively. We take advantage of the method in \cite{BelBarkai:2006} and find that for the PDF $f(\bar{p}_{\mathcal{D}})$, the dominant role is played by the distributions of the sojourn times in the two states, denoted as $\psi_+(\tau)$ and $\psi_-(\tau)$, respectively, for states in positive and negative half-spaces. As usual, we deal with the PDFs by using the techniques of Laplace transform, defined as
\begin{equation}
  \hat{\psi}_{\pm}(\lambda) = \int_0^{\infty} e^{-\lambda t}\psi_{\pm}(t) dt.
\end{equation}
According to the above discussions that the distribution of waiting times is space-dependent exponential form, we know that the mean of waiting time is finite and so is $\psi_{\pm}(t)$.
Therefore, the expansions of $\psi_{\pm}(t)$ for small $\lambda$ are $\hat{\psi}_{\pm}(\lambda) = 1- A_{\pm}\lambda$, where $A_{\pm}$ are their first moments, respectively.
For the occupation fraction $\bar{p}_+=T_+/t$, the result in \cite{BelBarkai:2006} is
\begin{equation}\label{R+}
  \mathcal{R}_+=\frac{A_+}{A_-}
\end{equation}
in \eqref{pdf4} and the PDF of $\bar{p}_+$ is
\begin{equation}\label{pdf5}
  f(\bar{p}_+)=\delta\left(\bar{p}_+-\frac{\mathcal{R}_+}{1+\mathcal{R}_+}\right),
\end{equation}
a $\delta$-function due to the finite mean of $\hat{\psi}_{\pm}(\lambda)$ (i.e., $\mu=1$ in \eqref{pdf4}).

\subsection{Calculations of $A_+$ and $A_-$}
The most important thing remaining is to calculate the two values $A_+$ and $A_-$, i.e., the means of the sojourn times $T_{\pm}$. 
 We will focus on the state in positive half-space while the case of the negative one is similar. The sojourn time $T_+$ in positive half-space can be viewed as the first-passage time of a particle starting at the position $\epsilon>0$ and reaching the origin $x=0$ for the first time. We eventually obtain the mean first-passage time $A_+$ by taking $\epsilon\rightarrow0$. Intuitively, the value $A_+$ must tend to zero when $\epsilon\rightarrow0$ since the particle has arrived at $x=0$ at initial time. This is completely right, but it could still provide us with a correct result of $\mathcal{R}_+$ in \eqref{R+}. This technique has also been used to calculate the area under the Brownian excursion \cite{MajumdarComtet:2005,Majumdar:2005} and that under the Bessel excursion \cite{BarkaiAghionKessler:2014,KesslerBarkai:2012}.

Firstly, let us pay attention to the PDF $P(x_0,x,t)$  of finding a particle at position $x$ at time $t$ starting from initial position $x_0$. It satisfies the backward Kolmogorov equation \cite{ArecchiPolitiUlivi:1982,Risken:1989},
\begin{equation}\label{BKE}
  \frac{\partial P(x_0,x,t)}{\partial t}= \mathcal{L}_{\textrm{FP}}^\ast~P(x_0,x,t),
\end{equation}
where $\mathcal{L}_{\textrm{FP}}^\ast$ is the adjoint operator of $\mathcal{L}_{\textrm{FP}}$ in \eqref{FFP}, defined as
\begin{equation}
  \mathcal{L}_{\textrm{FP}}^\ast = f(x_0)\frac{\partial}{\partial x_0} + \sqrt{D(x_0)}\frac{\partial}{\partial x_0}\sqrt{D(x_0)}\frac{\partial}{\partial x_0}.
\end{equation}
Since the first-passage time that a particle starts at $x_0=\epsilon$ and reaches $x=0$ for the first time is considering, for \eqref{BKE} we set the absorbing boundary condition at $x=0$. With this boundary condition, the survival probability $S(x_0,\tau_+)$, namely, the probability that a particle starts at $x_0$ and keeps in the positive half-space before the time $\tau_+$, satisfies \cite{Gardiner:1983,DengWuWang:2015}
\begin{equation}\label{SurvivalP}
  \textrm{Prob}\{t\geq \tau_+\}=S(x_0,\tau_+)=\int_0^{\infty}P(x_0,x,\tau_+)dx,
\end{equation}
where $t$ is the exact time that the particle leaves the positive half-space.
From the second equality in \eqref{SurvivalP}, we know that $S(x_0,\tau_+)$ also satisfies the backward Kolmogorov equation \eqref{BKE} with the same boundary condition.
From the first equality in \eqref{SurvivalP}, one can obtain the relation between the PDF of first-passage time $Q(x_0,\tau_+)$ and the survival probability
\begin{equation}
  Q(x_0,\tau_+)= -\frac{\partial}{\partial \tau_+}S(x_0,\tau_+);
\end{equation}
thus $Q(x_0,\tau_+)$ also satisfies the backward Kolmogorov equation \eqref{BKE}, i.e.,
\begin{equation}\label{EqQ}
  \frac{\partial Q(x_0,\tau_+)}{\partial \tau_+}= \mathcal{L}_{\textrm{FP}}^\ast~Q(x_0,\tau_+)
\end{equation}
with the normalization $\int_0^{\infty}Q(x_0,\tau_+)d\tau_+=1$.  Multiplying $\tau_+$ and integrating over $\tau_+$ in both sides of \eqref{EqQ}, we obtain the equation for the first moment of first-passage time $\langle\tau_+\rangle$:
\begin{equation}\label{EqMFPT}
  \mathcal{L}_{\textrm{FP}}^\ast~\langle\tau_+\rangle=-1.
\end{equation}
The boundary condition of \eqref{EqMFPT} is the same as those of $S(x_0,\tau_+)$ (i.e., absorbing at $x=0$), due to its relation with the survival probability $S(x_0,\tau_+)$:
\begin{equation}
  \langle\tau_+\rangle = \int_0^{\infty}\tau_+ Q(x_0,\tau_+)d\tau_+= \int_0^{\infty} S(x_0,\tau_+) d\tau_+.
\end{equation}
With the specified boundary condition, Eq. \eqref{EqMFPT} has a unique solution \cite{Gardiner:1983,ArecchiPolitiUlivi:1982}
\begin{equation}\label{solutionMFPT}
  \langle\tau_+\rangle (x_0) = \int_0^{x_0} W^{-1}(y)\int_y^{\infty}\frac{W(x)}{D(x)}dxdy,
\end{equation}
where $W(x)=\sqrt{D(x)}\exp[\int^x f(x')/D(x')]dx'$.

Now, we come back to the initial task of calculating $A_{\pm}$: the means of the sojourn time $\psi_{\pm}(t)$. For this, taking $x_0=\epsilon$ in \eqref{solutionMFPT} and letting $\epsilon\rightarrow0$ result in
\begin{equation}\label{A+}
  A_+=\lim_{\epsilon\rightarrow0} \langle\tau_+\rangle (\epsilon) \simeq \epsilon\cdot W^{-1}(0)\int_0^{\infty}\frac{W(x)}{D(x)}dx.
\end{equation}
On the other hand, we calculate $A_-$ by assuming that the particle starts at $x_0=-\epsilon$ with the boundary condition absorbing at $x=0$. The mean first-passage time $\langle\tau_-\rangle$ for a particle reaching $x=0$ for the first time also satisfies \eqref{EqMFPT} and its solution is
\begin{equation}
  \langle\tau_-\rangle (x_0) = \int_{x_0}^0 W^{-1}(y)\int_{-\infty}^y\frac{W(x)}{D(x)}dxdy.
\end{equation}
Being similar to \eqref{A+}, there exists
\begin{equation}\label{A-}
  A_-=\lim_{\epsilon\rightarrow0} \langle\tau_-\rangle (-\epsilon) \simeq \epsilon\cdot W^{-1}(0)\int_{-\infty}^0\frac{W(x)}{D(x)}dx,
\end{equation}
and then $\mathcal{R}_+$ in \eqref{R+} is obtained as
\begin{equation}\label{R+1}
\begin{split}
    \mathcal{R}_+ = \frac{\int_0^{\infty}\ W(x)/D(x) dx }{ \int_{-\infty}^0 W(x)/D(x) dx }
     = \frac{\int_0^{\infty}p^{st}(x)dx  }{  \int_{-\infty}^0p^{st}(x)dx   } ,
\end{split}
\end{equation}
where the steady solution is given in \eqref{pst0}. Though $A_{\pm}\rightarrow0$ as $\epsilon\rightarrow0$, the quantity of $\mathcal{R}_+$ of interest only depends on the speed of their tendencies to zero. Substituting this result into \eqref{pdf5} leads to the PDF of  the occupation fraction in positive half-space
\begin{equation}
    f(\bar{p}_+)
    =\delta(\bar{p}_+ - P_+^{st}),
\end{equation}
where $P_+^{st}=\int_0^{\infty} p^{st}(x)dx$ is the ensemble average of the occupation time in positive half-space in the steady state. This equation shows that the ergodic hypothesis is valid with respect to the occupation fraction in positive half-space.

More generally, for any domain $\mathcal{D}$, the result \eqref{pdf3} is also valid, such as a disconnected domain $\mathcal{D}=\{|x|>a\}$. In this case, the whole domain is divided into three sub-domains $\mathcal{D}_1=\{x<-a\}$, $\mathcal{D}_2=\{-a\leq x\leq a\}$, and $\mathcal{D}_3=\{x>a\}$. The sojourn time in these three sub-domains are denoted as $\psi_1(\tau),\, \psi_2(\tau), \,\psi_3(\tau)$ with finite means $A_1,\, A_2,\, A_3$, respectively. Following the above discussions, the key is to calculate $(A_1+A_3)/A_2$. Taking advantage of the result \eqref{R+1} and considering the domains $\mathcal{D}_1$ and $\mathcal{D}_2$ as a whole, we have
\begin{equation}\label{disconnect1}
  \frac{A_1+A_2}{A_3}=\frac{\int_{-\infty}^a p^{st}(x)dx}{\int_a^{\infty} p^{st}(x)dx};
\end{equation}
similarly, considering the domains $\mathcal{D}_2$ and $\mathcal{D}_3$ as a whole, there exists
\begin{equation}\label{disconnect2}
  \frac{A_1}{A_2+A_3}=\frac{\int_{-\infty}^{-a} p^{st}(x)dx}{\int_{-a}^{\infty} p^{st}(x)dx}.
\end{equation}
Combining \eqref{disconnect1} and \eqref{disconnect2}  results in
\begin{equation} \label{frac}
  \frac{A_1+A_3}{A_2}=\frac{\int_{-\infty}^{-a} p^{st}(x)dx+\int_a^{\infty} p^{st}(x)dx}{\int_{-a}^{a} p^{st}(x)dx}.
\end{equation}
Substituting \eqref{frac} into \eqref{pdf5} leads to the PDF of the occupation fraction $\bar{p}_\mathcal{D}$ in disconnected domains $\mathcal{D}_1$ and $\mathcal{D}_3$:
\begin{equation}
    f(\bar{p}_\mathcal{D})
    =\delta(\bar{p}_\mathcal{D} - P_{dis}^{st}),
\end{equation}
where $P_{dis}^{st}=\int_{x>a} p^{st}(x)dx$. So, it implies that the ergodic hypothesis is valid for the occupation fraction in any domain $\mathcal{D}$.

\subsection{Simulations}
\begin{figure}
  \centering
  \includegraphics[width=8cm,height=6cm]{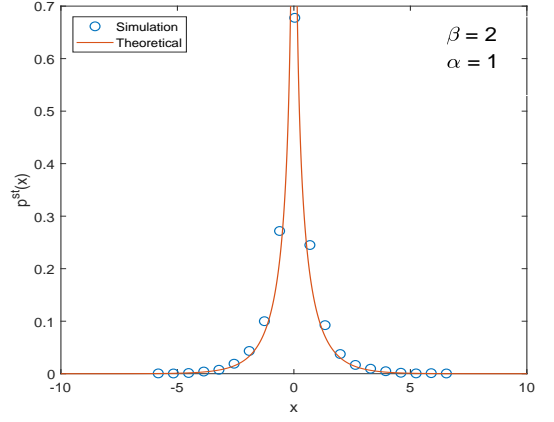}\\
  \caption{Steady solution $p^{st}(x)$ for $\beta=2$ and $\alpha=1$. We use trajectories up to $10^4$ and measurement time up to $T=10^3$. Other parameters: $U_0=1/2$, $D_0=1$, and $A_0=0.01$. The solid line denotes the theoretical result for long times and the markers are for the simulation results.}\label{fig1-1}
\end{figure}

\begin{figure}
  \centering
  \includegraphics[width=8cm,height=6cm]{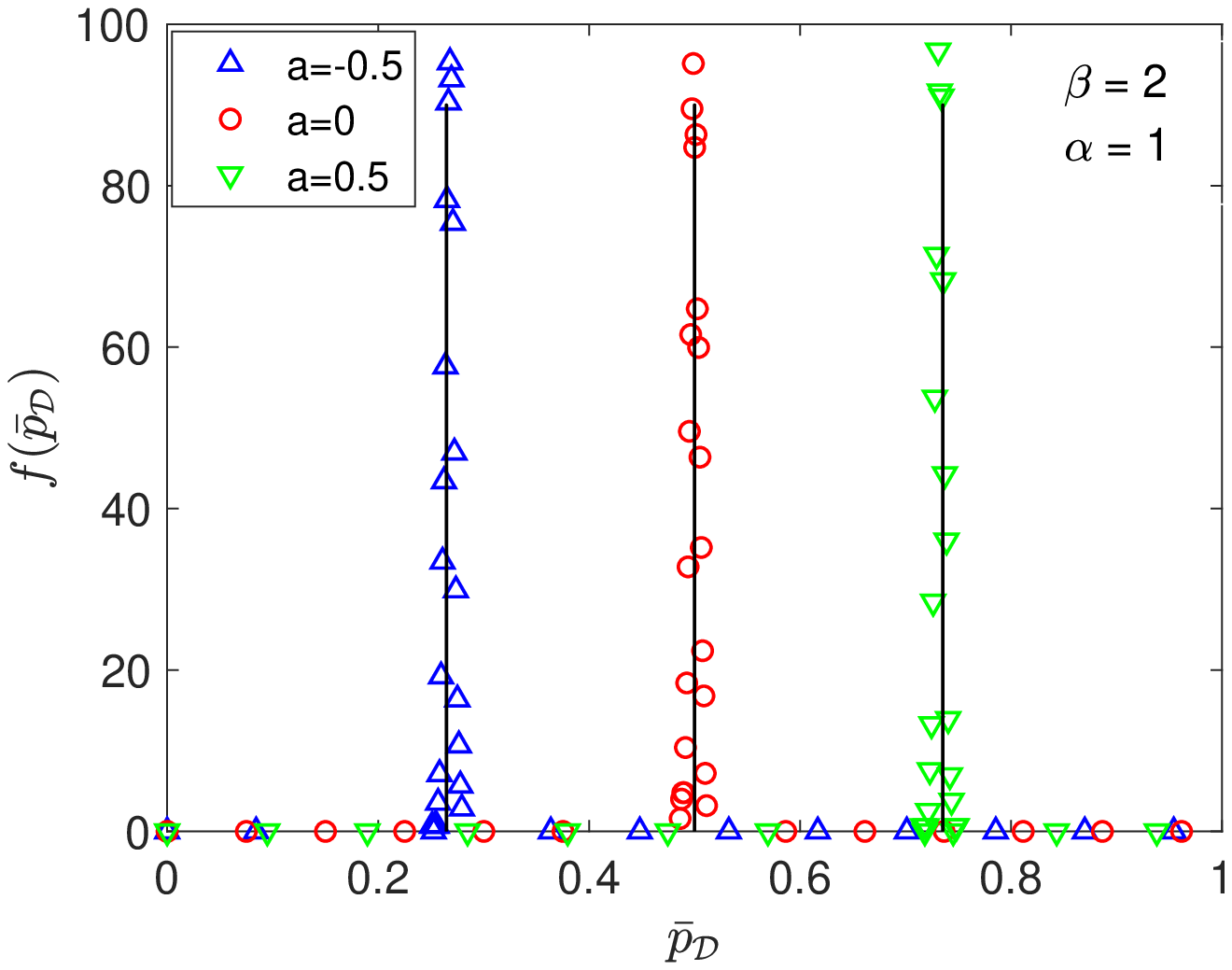}\\
  \caption{PDF $f(\bar{p}_{\mathcal{D}})$ of the occupation fraction in domain $\mathcal{D}=\{x<a\}$ with three different values $a=-0.5$, $0$, $0.5$, respectively. We use trajectories up to $10^3$ and measurement time up to $T=10^4$. Other parameters: $\beta=2$, $\alpha=1$, $U_0=1/2$, $D_0=1$, and $A_0=1$. The solid lines denote the theoretical results: $\delta$-function centered at three different values $\delta(\bar{p}_{\mathcal{D}}-0.2647)$, $\delta(\bar{p}_{\mathcal{D}}-0.5)$, and $\delta(\bar{p}_{\mathcal{D}}-0.7353)$ for long times
  and the markers are for the simulation results.}\label{fig1-2}
\end{figure}

\begin{figure}
  \centering
  \includegraphics[width=8cm,height=6cm]{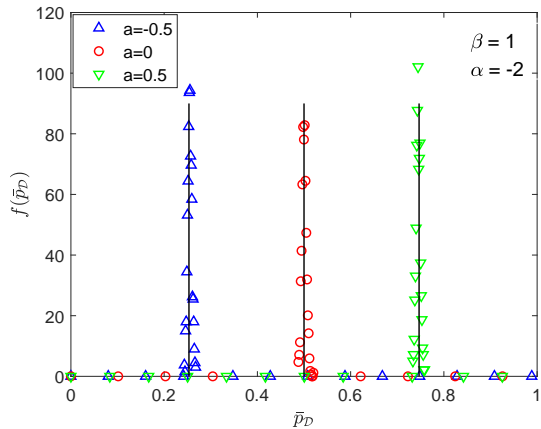}\\
  \caption{PDF $f(\bar{p}_{\mathcal{D}})$ of the occupation fraction in domain $\mathcal{D}=\{x<a\}$ with three different values $a=-0.5$, $0$, and $0.5$, respectively. We use trajectories up to $10^3$ and measurement time up to $T=10^4$. Other parameters: $\beta=1$, $\alpha=-2$, $U_0=1$, $D_0=1$, and $A_0=1$. The solid lines denote the theoretical results: $\delta$-function centered at three different values $\delta(\bar{p}_{\mathcal{D}}-0.2532)$, $\delta(\bar{p}_{\mathcal{D}}-0.5)$,  $\delta(\bar{p}_{\mathcal{D}}-0.7468)$ for long times
  and the markers are for the simulation results.}\label{fig1-3}
\end{figure}

We first simulate the steady solution $p^{st}(x)$ in \eqref{pst} for the case of $\beta>\alpha$. In Fig. \ref{fig1-1}, they are specified as $\beta=2$ and $\alpha=1$. In a regularized form, the space-dependent diffusivity and the potential well are $D(x)=|x|+A_0$ and $U(x)=(|x|+A_0)^2/2$, respectively, where $A_0=0.01$. Then the theoretical solution is $p^{st}(x)\propto (|x|+A_0)^{-1/2}\exp(-|x|)$, which is confirmed by the simulations in Fig. \ref{fig1-1} with trajectories up to $10^4$ and the measurement time up to $10^3$.
In Fig. \ref{fig1-2}, we present the PDFs of the occupation fraction in domain $\mathcal{D}=\{x<a\}$ with three different values $a=-0.5$, $0$, and $0.5$, respectively. In simulations, the same $D(x)$ and $U(x)$  are used as those in Fig. \ref{fig1-1}. We use trajectories up to $10^3$ and the measurement time up to  $10^4$. Similarly, we also simulate the PDFs of the occupation fraction for the case of $\alpha<0$ in Fig. \ref{fig1-3}. In this case, we take $D(x)=1/(1+|x|^{2})$ and $U(x)=|x|$ (i.e., $\beta=1$ and $\alpha=-2$). The simulations in Fig. \ref{fig1-3} are also consistent to the theoretical result \eqref{pdf3}. The space-dependent diffusivity $D(x)$ with $\alpha=-2$ is common in biological system, e.g., the small fluorescently labelled proteins in the cytoplasm of mammalian NLFK and HeLa cells \cite{Kuhn.etal:2011}.

It deserves to be mentioned that the exponent $\alpha$ can be bigger than $2$ when reflecting boundary condition is applied. Since, in some sense, the reflecting boundary condition at $\pm L$ can be regarded as the potential well $U(x)=|x/L|^\beta/\beta$ with $\beta\rightarrow\infty$; for any given $\alpha$, the condition $\beta>\alpha$ is always satisfied and the result of the PDF of the occupation fraction is also valid. In this case, the steady solution is $p^{st}(x)\propto |x|^{-\alpha/2}$ in \eqref{pst}, which only depends on the diffusivity $D(x)$. The steady solution in bounded domain varies for different processes. Some interesting results have been found for L\'{e}vy flight and L\'{e}vy walk \cite{DybiecGudowskaBarkaiDubkov:2017}, fractional Brownian motion \cite{WadaVojta:2018} in a bounded domain.

One specific example for the process driven by a Langevin equation with multiplicative noise in a bounded domain has been discussed in \cite{WangChenDeng:2018}, where the PDF of the occupation fraction in positive half-space was derived to be a $\delta$-function based on the backward Feynman-Kac equation. The multiplicative noise in \cite{WangChenDeng:2018} was interpreted in the It\^{o} sense, different from the Stratonovich sense considered in this paper. The results in this paper can be easily extended to other different interpretations of the multiplicative noises, including It\^{o} sense \cite{Ito:1950} and H\"{a}nggi-Klimontovich sense \cite{Hanggi:1982,Klimontovich:1990}. For different interpretations, one just needs to modify the external force term in the Fokker-Planck equation \eqref{FFP} and obtain a similar steady solution $p^{st}(x)$, which will not bring any difficulties for the discussions in this paper.

\section{Infinite ergodic theory for $\beta=\alpha$}\label{Sec4}
Now, we turn to the case of $\beta=\alpha>0$, where the steady solution obtained in \eqref{pst} is
\begin{equation}\label{pst1}
  p^{st}(x)=\frac{N}{\sqrt{D_0}}|x|^{-\alpha/2-\alpha U_0/D_0}.
\end{equation}
It can be seen that this PDF decays as power-law and its tail now relies heavily on the magnitudes $D_0$ in the diffusivity $D(x)$ and $U_0$ in the potential well $U(x)$. The PDF \eqref{pst1} can be normalized only when $\alpha/2+\alpha U_0/D_0>1$; under this condition, the ensemble average of the observable occupation fraction in the positive half-space is finite, namely, $\int_0^\infty p^{st}(x)dx<\infty$  (note that $p^{st}(x)$ is regular at origin as mentioned in Sec. \ref{Sec2}). But if the observable of interest is second moment, it is only finite under a stronger request: $\alpha/2+\alpha U_0/D_0>3$, which shows that a relatively stronger external potential well or a larger radio $U_0/D_0$ is responsible for a finite higher-order moment with respect to the steady solution. What happens when these requests are not satisfied? As we all know, the higher-order moments mainly depend on the tail of the PDF. But the steady solution $p^{st}(x)$ do not provide a quite correct information at its tail when we measure the higher-order moments. In turn, $p^{st}(x)$ is correct enough to calculate some suitable statistics. For example, we can get the correct ensemble averaged occupation fraction rather than second moment under the condition $1<\alpha/2+\alpha U_0/D_0<3$.

What about the ensemble averaged occupation fraction under the condition $\alpha/2+\alpha U_0/D_0<1$? Since $p^{st}(x)$ fails to describe the information at its tail, instead, we consider a time-dependent solution to obtain a correct decay rate. The common methods to obtain the solution are the eigenfunction expansion \cite{Risken:1989,MetzlerBarkaiKlafter:1999} and scaling ansatz \cite{KesslerBarkai:2010,BarkaiAghionKessler:2014} of the Fokker-Planck equation \eqref{FFP}. We will apply the latter to solve \eqref{FFP} with the general $U(x)$ and $D(x)$. Due to the complexity of the system for any $\alpha<2$ (when $\alpha=2$, $\alpha/2+\alpha U_0/D_0<1$ does not hold), we will simplify the problem firstly through the variable substitution \eqref{substitution}. Then, Eq. \eqref{model2} reduces to the Langevin equation of $y(t)$:
\begin{equation}\label{Eq_y}
  \frac{dy}{dt}= -\frac{A}{y} + \xi(t),
\end{equation}
where
\begin{equation}\label{CondA}
  A=\frac{\alpha}{2-\alpha}\frac{U_0}{D_0}< \frac{1}{2}.
\end{equation}
This system describes the motion of a Brownian particle in a logarithmic potential, which has been discussed in detail in \cite{Lutz:2004,DechantLutzKesslerBarkai:2011}. The solution $y(t)$ in \eqref{Eq_y} is also named as Bessel process, being related to a free Brownian particle starting from the origin in high dimensions.
The exact form of its PDF $p(y,t)$ can be found in
\cite{Bray:2000,MartinBehnGermano:2011,DechantLutzBarkaiKessler:2011,LeibovichBarkai:2018}. On the other hand, to avoid complicated calculations, it is easy to do scaling ansatz for
Eq. \eqref{Eq_y}, which has been mentioned in \cite{KesslerBarkai:2010}, and we present the details in Appendix \ref{APPb} for completeness. For $|y|\ll t^{1/2}$, we obtain the asymptotic behavior of the solution $p(y,t)$ in \eqref{Appb5}
\begin{equation}
  p(y,t)\simeq N_g t^{A-\frac{1}{2}} |y|^{-2A}
\end{equation}
with the normalization parameter $N_g=2^{A-1/2}[\Gamma(1/2-A)]^{-1}$.
Considering the relation between $y(t)$ and $x(t)$ in \eqref{substitution}, i.e., $|y|= \frac{\sqrt{2/D_0}}{2-\alpha}|x|^{\frac{2-\alpha}{2}}$, we obtain the asymptotic behavior of the solution $p(x,t)$: for $|x|\ll t^{1/(2-\alpha)}$,
\begin{equation}\label{pxtalpbet}
\begin{split}
    p(x,t)&\simeq N_x t^{A-\frac{1}{2}} |x|^{-A(2-\alpha)-\alpha/2},  \\
\end{split}
\end{equation}
where $N_x=(2-\alpha)^{2A}D_0^{A-1/2}[2\Gamma(1/2-A)]^{-1}$. Note that the exponent of $x$ is consistent with the steady solution $p^{st}(x)$ in \eqref{pst1}, which shows the potential value of the non-normalized steady solution. This consistency also explains why we consider the asymptotic behavior in $|x|\ll t^{1/(2-\alpha)}$ (i.e., the small $z$ in scaling ansatz in Appendix \ref{APPb}).
Moreover, there is a relation between them
\begin{equation}\label{Relation}
  \lim_{t\rightarrow\infty} p(x,t) t^{\frac{1}{2}-A} =  p^{st}(x).
\end{equation}
From \eqref{Relation}, the parameter $N$ in \eqref{pst1} can be determined as $N=(2-\alpha)^{2A}D_0^{A}[2\Gamma(1/2-A)]^{-1}$. It looks natural that $p^{st}(x)$ is non-normalized, since the term $t^{\frac{1}{2}-A}$ tends to infinity for $A<1/2$. On the other hand, though $p^{st}(x)$ is non-normalized, it is still useful since it only differs from the real $p(x,t)$ by a known term of $t$.
With this relation, the infinite-ergodic theory can be built up, uncovering the relation between the time average and ensemble average of an observable.

Considering an observable $\mathcal{O}(x)$, its ensemble average is defined as
\begin{equation}
  \langle \mathcal{O}(x)\rangle = \int_{-\infty}^{\infty} \mathcal{O}(x) p(x,t) dx,
\end{equation}
which can be represented through the non-normalized steady solution in \eqref{Relation} for long times
\begin{equation}\label{EA1}
  \lim_{t\rightarrow\infty} \langle \mathcal{O}(x)\rangle =   t^{A-\frac{1}{2}} \int_{-\infty}^{\infty} \mathcal{O}(x) p^{st}(x) dx,
\end{equation}
 if the integral on the right-hand side exists. This condition is not rigid for many observables, especially for those decaying for large $x$ which cure the non-integrability of $p^{st}(x)$ at infinity.
It also confirms the explanation that the steady solution $p^{st}(x)$ is valid except its tail-part.
A common one is the pulse function $\Theta_\mathcal{D}(x)$ with respect to a finite interval $\mathcal{D}$ (e.g., $\mathcal{D}=[-a,a]$ in the following), which is defined as $\Theta_\mathcal{D}(x)=1$ for $x\in\mathcal{D}$ and $\Theta_\mathcal{D}(x)=0$ otherwise. So it vanishes at infinity and it is integrable with respect to $p^{st}(x)$.

Now, we turn to the time average of the observable $\mathcal{O}(x)$, which is defined as
\begin{equation}
  \overline{\mathcal{O}(x)} = \frac{1}{t}\int_0^t \mathcal{O}(x(t')) dt';
\end{equation}
and the corresponding ensemble-time average is
\begin{equation}
    \langle \overline{\mathcal{O}(x)} \rangle = \frac{1}{t}\int_0^t \langle\mathcal{O}(x(t'))\rangle dt'.
\end{equation}
Taking the limit $t\rightarrow \infty$ and using Eq. \eqref{EA1}, we obtain
\begin{equation}\label{IET}
  \lim_{t\rightarrow\infty} \langle \overline{\mathcal{O}(x)} \rangle / \langle \mathcal{O}(x)\rangle = \frac{2}{2A+1} ,
\end{equation}
where the constant factor $2/(2A+1)$ comes from the time-integral of the term $t^{A-1/2}$ in \eqref{EA1}.
Compared to the result \eqref{pdf3} in the case of $\beta>\alpha$, where the normalized steady solution $p^{st}(x)$ directly decides the equivalent time and ensemble average,
Eq. \eqref{IET} shows that time and ensemble averages are still related but only differ by a constant factor for long times.
In addition, the two kinds of averages are both found to be dependent on the non-normalized steady state $p^{st}(x)$ and decay as $t^{A-\frac{1}{2}}$, which can be seen from equations \eqref{EA1} and \eqref{IET}. It means that the number of the particles in the finite domain $\mathcal{D}$ decays as $t^{A-\frac{1}{2}}$ and the diffusion behavior goes on even for long times. In the case of $A<1/2$, i.e., $U_0/D_0<(2-\alpha)/(2\alpha)$, the potential well is too weak, compared with the space-dependent diffusivity, to suppress the diffusion; more and more particles go to infinity and thus they are not counted by the observable $\mathcal{O}(x)$.

\subsection{PDF of the time averaged observable}
The distribution of the time averaged occupation fraction $f(\bar{p}_{\mathcal{D}})$ has been shown to be a $\delta$-function in \eqref{pdf3}, implying the ergodic phase for $\beta>\alpha$; on the contrary, the distribution for the case $\beta=\alpha$ becomes different, which will be shown to be broad and asymmetric in the following. The scatter of the amplitude of the time average for a set of individual trajectories is a useful indicator for the degree of non-ergodicity and the classification of different processes \cite{MetzlerJeonCherstvyBarkai:2014}. We are interested in the distribution of the time averaged observable $\overline{\mathcal{O}(x)}$ around its mean for long times, so we define the random variable of dimensionless
\begin{equation}\label{eta}
  \eta=\frac{\overline{\mathcal{O}(x)}}{\langle \overline{\mathcal{O}(x)} \rangle}.
\end{equation}
Especially, we consider the observable $\mathcal{O}(x)$ to be the pulse function $\Theta_\mathcal{D}(x)$ with $\mathcal{D}=[-a,a]$. Then the time averaged observable $\overline{\mathcal{O}(x)}$ becomes the occupation fraction $p_{\mathcal{D}}$ in domain $\mathcal{D}$.

\begin{figure}
  \centering
  \includegraphics[scale=0.5]{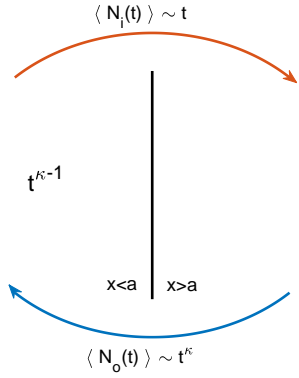}\\
  \caption{Simple sketch map for the particles walking in two states. The black vertical line is the division line of the two states: inside or outside the domain $\mathcal{D}=[-a,a]$. For simplicity, we only show the positive half-space here due to symmetry. The left side ($x<a$) denotes the state inside the domain $\mathcal{D}$ while the right side ($x>a$) is for another state outside $\mathcal{D}$. The maroon arrow on the top denotes that the particles change from state inside $\mathcal{D}$ to state outside $\mathcal{D}$, with the average number $\langle N_i(t)\rangle \sim t$. The blue arrow on the bottom signifies that the particles change from state outside $\mathcal{D}$ to state inside $\mathcal{D}$, with the average number $\langle N_o(t)\rangle \sim t^\kappa$. Therefore, the total number in domain $\mathcal{D}$ decays as $t^{\kappa-1}$.
  }\label{fig2-1}
\end{figure}

When $\beta=\alpha$ and $A<1/2$ in \eqref{CondA}, the diffusion of the particles cannot be controlled by the potential well. More and more particles leave the bounded domain $\mathcal{D}$ and never come back, resulting in the decay of the number of the particles in $\mathcal{D}$ as $t^{A-\frac{1}{2}}$. The trajectory of a particle can be described by two states: leaving the domain $\mathcal{D}$ and returning to the domain. We show a simple sketch map in Fig. \ref{fig2-1}, where the negative half-space is omitted due to symmetry.
The two states $x<a$ and $x>a$ have completely different properties. The mean of the sojourn time in the bounded domain $\mathcal{D}$ (on the left side $x<a$) is finite, being similar to the case of $\beta>\alpha$. But for the sojourn time outside the domain $\mathcal{D}$ (on the right side $x>a$), its PDF is heavy-tailed $\phi(\tau)\sim \tau^{-1-\kappa},\,0<\kappa<1$ for large $\tau$ with infinite mean value due to a large fluctuation.
The events that the particle outside (inside) $\mathcal{D}$ crosses the division point $x=a$ are recurrent. These events are not correlated and the PDF of the number of renewals $N_o(t)\,(N_i(t))$ before time $t$ can be solved by renewal theory \cite{Cox:1962}. Importantly, the occupation time $T_{\mathcal{D}}(t)$ inside $\mathcal{D}$ is proportional to $N_o(t)$ due to the finite mean of the sojourn time inside $\mathcal{D}$. Therefore, the distribution of the random variable $\eta$ is related to the distribution of $N_o(t)$:
\begin{equation}
  \eta= \frac{p_{\mathcal{D}}}{\langle p_{\mathcal{D}}\rangle}  = \frac{T_{\mathcal{D}}(t)}{\langle T_{\mathcal{D}}(t)\rangle}
  =  \frac{N_o(t)}{\langle N_o(t) \rangle }.
\end{equation}

Now only the exponent $\kappa$ in the heavy-tailed law $\phi(\tau)$ remains to be determined to obtain the PDF of $N_o(t)/\langle N_o(t) \rangle$. Actually, it only depends on the parameter $A$. Since the mean of the sojourn time inside $\mathcal{D}$ is finite, the average number of renewals is $\langle N_i(t)\rangle\propto t$. Similarly, considering the PDF of the sojourn time outside the domain is heavy-tailed, we have $\langle N_o(t)\rangle\propto t^\kappa$ \cite{Cox:1962,GodrecheLuck:2001,WangSchulzDengBarkai:2018}. Then the net number of the particles inside $\mathcal{D}$ decays as $t^{\kappa-1}$, which should be consistent to the result previously obtained $t^{A-1/2}$, and thus it leads to
\begin{equation}\label{kappa}
  \kappa=A+1/2.
\end{equation}
Then the PDF of $\eta$ is known as the Mittag-Leffler distribution \cite{Aaronson:1997,Feller:1971}:
\begin{equation}\label{MLdist}
  f(\eta)= \frac{\Gamma^{1/\kappa}(1+\kappa)}{\kappa \eta^{1+1/\kappa}} L_\kappa \left[ \frac{\Gamma^{1/\kappa}(1+\kappa)}{\eta^{1/\kappa}} \right],
\end{equation}
where $L_\kappa[\cdot]$ is the one-sided L\'{e}vy density with order $\kappa$. Note that $f(\eta)$ in \eqref{MLdist} is obtained for large $t$ and it finally becomes independent on $t$. Though the total time $t$ in the renewal process consists of two parts: the time particles spend inside and outside $\mathcal{D}$, the first part inside $\mathcal{D}$ can be statistically omitted compared to the second part.

We have used $A$ to characterize the radio between the strength of the potential well $U(x)$ and space-dependent diffusivity $D(x)$ in \eqref{CondA}, and then $A$ determines the parameter $\kappa$ in \eqref{kappa} and the distribution of the time average $f(\eta)$. It is known that $f(\eta)\rightarrow \delta(\eta-1)$, returning to the ergodic phase when $\kappa\rightarrow1$. On the contrary, when $\kappa$ takes its minimum value $1/2$, the distribution $f(\eta)$ becomes the broadest, implying the most significant non-ergodic behavior.  Therefore, for the value of $A$ turning from $0$ to $1/2$, $\kappa$ is from $1/2$ to $1$, and the system changes from non-ergodicity to the ergodic phrase.

\subsection{Simulations}

\begin{figure}
  \centering
  \includegraphics[scale=0.5]{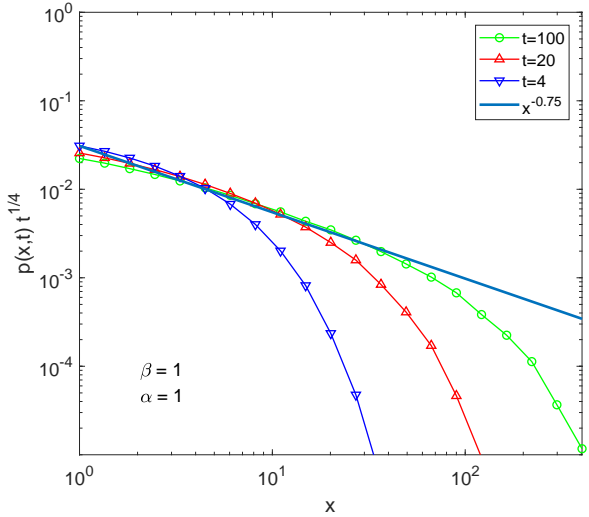}\\
  \caption{Scaled PDF $p(x,t)t^{1/4}$ versus $x$ in the positive half-space at three different times $t=4$, $t=20$, and $t=100$. $10^6$ trajectories are used. Other parameters are $A_0=1$, $D_0=1$, and $U_0=1/4$. The solid line is the theoretical result, i.e., the non-normalized steady solution $p^{st}(x)\propto x^{-3/4}$ for large $x$ in \eqref{pst1}. The markers are the simulation results. As time goes on, the PDF of simulation approaches the non-normalized solution $p^{st}(x)$.}\label{fig2-2}
\end{figure}

\begin{figure}
  \centering
  \includegraphics[scale=0.58]{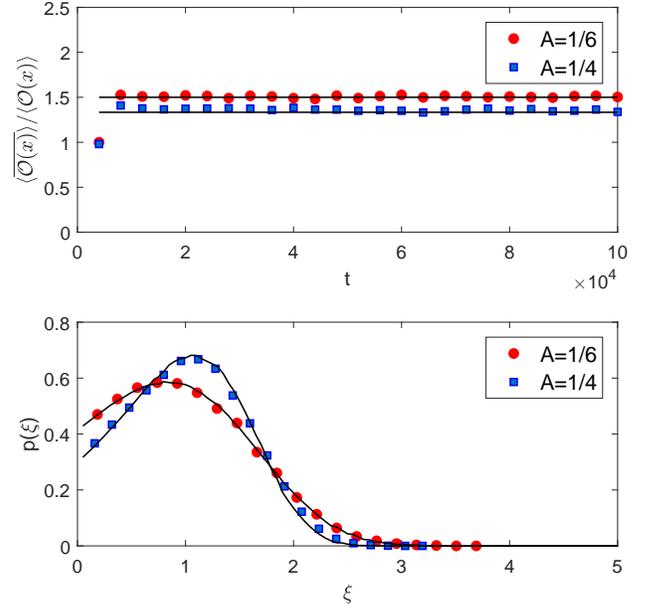}\\
  \caption{Upper panel: The radio $\langle \overline{\mathcal{O}(x)} \rangle / \langle \mathcal{O}(x)\rangle$ versus time $t$ for $\beta=\alpha=1$. The total measurement time is $T=10^5$ and the number of trajectories is $10^6$. Other parameters are $a=0.5$, $A_0=1$, $D_0=1$, $U_0=1/6$ for red circle, and $U_0=1/4$ for blue square. The black solid lines are the theoretical results in \eqref{IET}.
   Lower panel: The distribution of the time averages $\eta$ in \eqref{eta} for $\beta=\alpha=1$. The total measurement time is $T=10^4$ and the number of trajectories is $10^6$. Other parameters are $a=0.5$, $A_0=1$, $D_0=1$, $U_0=1/6$ for red circle, and $U_0=1/4$ for blue square. The black solid lines are the  Mittag-Leffler distribution \eqref{MLdist}.
   }\label{fig2-3}
\end{figure}

We first simulate the solution $p(x,t)$ in Fig. \ref{fig2-2} for the case of $\beta=\alpha=1$. The regularized factor is taken to be $A_0=1$ and other parameters are specified as $D_0=1,U_0=1/4$, and then $A=1/4$ in \eqref{CondA}.
The theoretical solution in \eqref{pxtalpbet} is $p(x,t)\simeq N_x t^{-1/4}|x|^{-3/4}$ for $|x|\ll t$. The auxiliary solid line in Fig. \ref{fig2-2} has the same slope as theoretical result $N_x|x|^{-3/4}$, which is also the non-normalized steady solution $p^{st}(x)$ in \eqref{pst1}. The prefactor of the auxiliary solid line is a little different from $N_x$ due to the regularized factor $A_0$. As the time goes by, the non-normalized solution $p^{st}(x)$ (solid line) is approached.

In the upper panel of Fig. \ref{fig2-3}, we demonstrate the result of the radio $\langle \overline{\mathcal{O}(x)} \rangle / \langle \mathcal{O}(x)\rangle$ in \eqref{IET}; due to its dependence on $A$, we choose parameters $\alpha=\beta=1$, $A_0=1$, $D_0=1$, and different $U_0$ ($U_0=1/6,\,1/4$) corresponding to different $A=1/6,\,1/4$ and different theoretical radio $\langle \overline{\mathcal{O}(x)} \rangle / \langle \mathcal{O}(x)\rangle=3/2,\,4/3$ for long times. The observed domain is $\mathcal{D}=[-0.5,\,0.5]$, i.e., $a=0.5$. The agreement between theoretical and simulated results can be seen in the upper panel with the measurement time up to $10^5$. In the lower panel, we verify the PDF $f(\eta)$ of the time averaged occupation time $\eta$ with the same parameters as above. So the theoretical PDF is the Mittag-Leffler distribution with order $\kappa=2/3$ and $3/4$ in \eqref{MLdist}, respectively, which is shown as the solid line and consistent to the simulation results in the lower panel. This theoretical lines can be obtained by numerical inverse Laplace transform, since the one-sided L\'{e}vy density is $L_\kappa(z)=\mathcal{L}_{s\rightarrow z}^{-1}[\exp(-s^\kappa)]$.

\section{Infinite ergodic theory for $\beta<\alpha$}\label{Sec5}
In the case of $\beta<\alpha$ and $\alpha>0$, the steady solution has been given in \eqref{pst} as
\begin{equation}\label{pst2}
  p^{st}(x)= \frac{N}{\sqrt{D_0}}|x|^{-\alpha/2} \exp\left(\frac{U_0\beta}{D_0(\alpha-\beta)}|x|^{-(\alpha-\beta)}\right),
\end{equation}
which can not be non-normalized for any parameters due to its large-$x$ behavior $|x|^{-\alpha/2}$. Similar to the case of $\beta=\alpha$, we need to solve the Fokker-Planck equation \eqref{FFP} to obtain a time-dependent solution. Since the potential well $U(x)$ in this case is too weak to change the shape of the solution $p(x,t)$ in \eqref{pdf1} for a free particle, we employ a different scaling ansatz for the time-dependent solution, and obtain
\begin{equation}\label{Relation2}
\begin{split}
    \lim_{t\rightarrow\infty}p(x,t)&=\frac{1}{\sqrt{4\pi D_0t}} \exp\left(-\frac{|x|^{2-\alpha}}{(2-\alpha)^2D_0t}\right) \cdot p^{st}(x)   \\[3pt]
    &\approx  \mathcal{Z}(t)\,  |x|^{-\alpha/2},
\end{split}
\end{equation}
for long times, where $\mathcal{Z}(t)=1/\sqrt{4\pi D_0t}$.  The details of the derivation of \eqref{Relation2} and $N=\sqrt{D_0}$ in \eqref{pst2} are provided in Appendix \ref{APPc}.
Similar to \eqref{Relation}, Eq. \eqref{Relation2} shows the relation between the non-normalized steady solution and the time-dependent solution for long times. That the term $\mathcal{Z}(t)$ decays as $t^{-1/2}$ shows the particles tend to accumulate at infinity. For some observable $\mathcal{O}(x)$, the relation between ensemble-time average and ensemble average of the observable can be obtained
\begin{equation}\label{IET2}
  \lim_{t\rightarrow\infty} \langle \overline{\mathcal{O}(x)} \rangle / \langle \mathcal{O}(x)\rangle = 2 ,
\end{equation}
and they both decay as $t^{-1/2}$. This result is consistent to the case of $A=0$ ($\kappa=1/2$) in the previous section $\beta=\alpha$, which means that the minimum value of $\kappa$ is $1/2$ and the non-ergodic behavior will not be stronger than the case of $\kappa=1/2$. This specific value $1/2$ is associated with the Gaussian white noise $\xi(t)$. More precisely, the diffusion behavior of a free Brownian particle is characterized by the scaling $x\sim t^{1/2}$, implying that the number of the particles in a bounded domain decays as $t^{-1/2}$. Even in a heterogeneous media, where the particle undergoes subdiffusion or superdiffusion, the asymptotic behavior of its PDF for large $t$ is also $t^{-1/2}$ in \eqref{pdf1} and \eqref{pdf2}.
Similarly, we define the random variable $\eta$ to characterize the distribution of the time averaged observable, and obtain that its PDF $f(\eta)$ also obeys Mittag-Leffler distribution with the order $\kappa=1/2$.

\subsection{Simulations}
\begin{figure}
  \centering
  \includegraphics[scale=0.5]{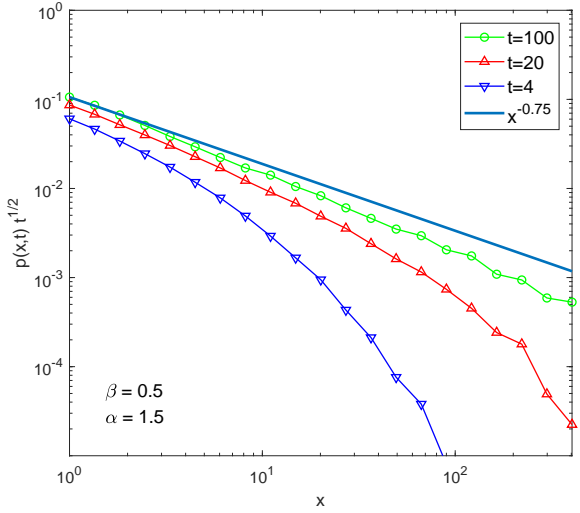}\\
  \caption{Scaled PDF $p(x,t)t^{1/2}$ versus $x$ in the positive half-space at three different times $t=4$, $t=20$, and $t=100$. $10^6$ trajectories are used. Other parameters are $A_0=1$, $D_0=1$, and $U_0=2$. The solid line is the theoretical result, i.e., the non-normalized steady solution $p^{st}(x)\propto x^{-3/4}$ for large $x$ in \eqref{pst2}. The markers are the simulation results. As the time goes on, the PDF of simulation approaches the non-normalized solution $p^{st}(x)$.}\label{fig3-1}
\end{figure}

\begin{figure}
  \centering
  \includegraphics[scale=0.5]{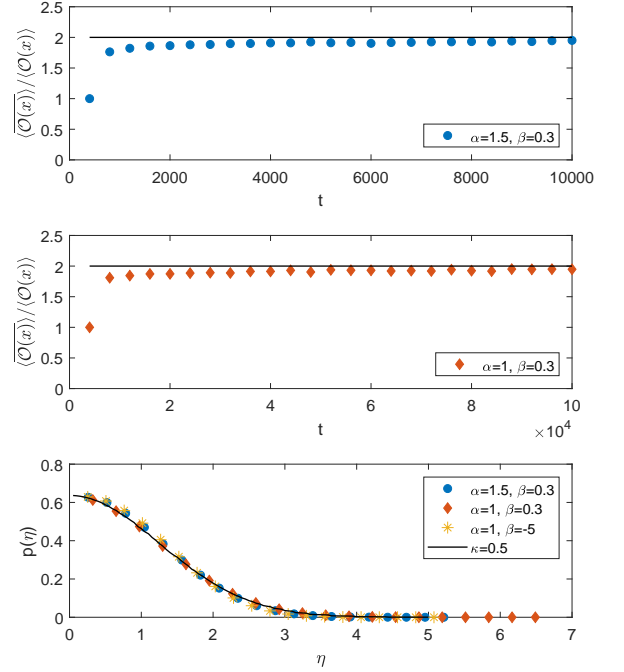}\\
  \caption{Upper panel: The radio $\langle \overline{\mathcal{O}(x)} \rangle / \langle \mathcal{O}(x)\rangle$ versus time $t$ for $\alpha=1.5$ and $\beta=0.3$. The total measurement time is $T=10^4$ and the number of trajectories is $10^6$. Other parameters are $a=0.5$, $A_0=1$, $D_0=1$, and $U_0=1$. The black solid lines are the theoretical results in \eqref{IET}.
  Middle panel: The radio $\langle \overline{\mathcal{O}(x)} \rangle / \langle \mathcal{O}(x)\rangle$ versus time $t$ for $\alpha=1$ and $\beta=0.3$. The total measurement time is $T=10^5$ and the number of trajectories is $10^6$. Other parameters are $a=0.5$, $A_0=1$, $D_0=1$, and $U_0=1$. The black solid lines are the theoretical results in \eqref{IET}. Comparing the upper and middle panels, a larger measurement time is needed for this system to reach the theoretical result \eqref{IET} with a smaller $\alpha$.
  Lower panel: The distribution of the time averages $\eta$ in \eqref{eta} for three sets of different parameters: $\alpha=1.5,\beta=0.3$; $\alpha=1,\beta=0.3$; and $\alpha=1,\beta=-5$. The total measurement time is $T=10^4$ and the number of trajectories is $10^6$. Other parameters are $a=0.5$, $A_0=1$, $D_0=1$, and $U_0=1$. The black solid lines are the Mittag-Leffler distribution with $\kappa=1/2$ in \eqref{MLdist}. The simulation results agree well with the theoretical results, even for $\beta<0$.
  }\label{fig3-2}
\end{figure}

We first simulate the solution $p(x,t)$ in Fig. \ref{fig3-1} for the case of $\alpha=1.5$ and $\beta=0.5$ with other parameters specified as $D_0=1$ and $U_0=2$.
The theoretical solution is the non-normalized steady solution in \eqref{pst2} $p^{st}(x)\propto x^{-3/4}$ for large $x$, shown as the solid line in Fig. \ref{fig3-1}.
As the time goes by, the non-normalized solution $p^{st}(x)$ (solid line) is approached.

In the upper and middle panels of Fig. \ref{fig3-2}, we demonstrate the results of the radio $\langle \overline{\mathcal{O}(x)} \rangle / \langle \mathcal{O}(x)\rangle$ in \eqref{IET} with two sets of different parameters: $\alpha=1.5,\,\beta=0.3$; and $\alpha=1,\,\beta=0.3$.
Other parameters are $D_0=U_0=1$. The observed domain is $\mathcal{D}=[-0.5,0.5]$. The agreement between theoretical and simulated results can be seen for total measurement time up to $10^4$ in the upper panel and $10^5$ in the middle panel. It implies that for stronger diffusivity $D(x)$ (larger $\alpha$), the system displays a more enhanced diffusion behavior and reaches the state \eqref{IET} using less time.
In the lower panel, we verify the PDF $f(\eta)$ of the time averaged occupation time $\eta$ with the same parameters as above and a set of new parameter with negative $\beta$: $\alpha=1,\beta=-5$. For the three sets of different parameters, all the theoretical PDFs are the Mittag-Leffler distribution with order $\kappa=1/2$ in \eqref{MLdist}, which is shown as the solid line, being consistent to the simulation results in the lower panel.

\section{Summary and conclusions}\label{Sec6}

\begin{figure}
  \centering
  \includegraphics[scale=0.5]{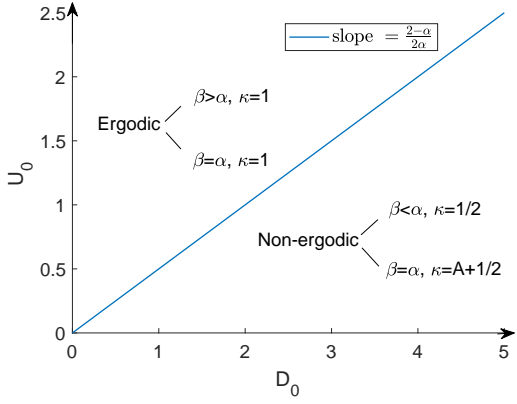}\\
  \caption{Ergodic properties: $U_0$ versus $D_0$ in the case of $\beta=\alpha$. The slope of the dividing line is $(2-\alpha)/(2\alpha)$. In the upper domain of this line, the system is ergodic with $\kappa=1$ in the distribution of the time averaged occupation time. But in the lower part of this line, this system is non-ergodic with $\kappa=A+1/2$. Other two cases ($\beta>\alpha$ and $\beta<\alpha$) belong to ergodic phase and non-ergodic phase with $\kappa=1$ and $\kappa=1/2$, respectively, for any positive $U_0$ and $D_0$.}\label{fig4}
\end{figure}

As a conclusion, we condense the main results for three different cases into a diagram presented in Fig. \ref{fig4}. The horizontal and vertical axes are the parameters $D_0$ and $U_0$ in diffusivity $D(x)$ and potential well $U(x)$, respectively. It has been fully discussed that the competition role played by $D_0$ and $U_0$ for $\alpha=\beta$ decides whether the system is ergodic or not in terms of the observable---occupation time in Sec. \ref{Sec4}. It can be noted that the dividing line shown in Fig. \ref{fig4} is
\begin{equation}
  \frac{U_0}{D_0}=\frac{2-\alpha}{2\alpha}
\end{equation}
from the condition $A<1/2$ in \eqref{CondA}. In the case of $\alpha=\beta$, the system is ergodic when $U_0/D_0$ is larger than the critical value $(2-\alpha)/(2\alpha)$ and non-ergodic otherwise. The distribution of the time averaged observable is a $\delta$-function in ergodic phase, i.e., $\kappa=1$ in the Mittag-Leffler distribution. In non-ergodic phase, the time average is a random variable and its distribution is the Mittag-Leffler distribution with order $\kappa$ depending on $U_0$, $D_0$, and $\alpha$, i.e., $\kappa=A+1/2$ with $A$ defined in \eqref{CondA}. Compared to $\beta=\alpha$, the results are relatively simple for another two cases $\beta>\alpha$ and $\beta<\alpha$ in Sec. \ref{Sec3} and Sec. \ref{Sec5}, respectively. For any positive $U_0$ and $D_0$, the former is ergodic with $\kappa=1$ and the latter is non-ergodic with $\kappa=1/2$ as shown in Fig. \ref{fig4}.

For the case of $\beta=\alpha$, there is an interesting phenomenon with the dividing line in Fig. \ref{fig4}. Since $\alpha>0$ has been assumed in this case, we consider two extreme values  $0$ and $2$ for $\alpha$. If $\beta=\alpha\rightarrow2$, the dividing line with slope $(2-\alpha)/(2\alpha)$ approaches the horizontal line ($D_0$-axis), implying the ergodic phase for any value of $U_0\neq0$. In contrast, for $\beta=\alpha\rightarrow0$, the dividing line approaches the vertical line ($U_0$-axis), implying the non-ergodic phase for any value of $D_0\neq0$. One example of the latter is a free Brownian particle exhibiting normal diffusion.

We take the potential well $U(x)$ and space-dependent diffusivity $D(x)$ to be the power-law form, due to its simple way to analyze the large-$x$ behavior. For general $U(x)$ and $D(x)$, e.g., the asymmetric ones and the bistable potential, the methods used and the results obtained in this paper are still valid. We find that a steady state can be reached when the potential well is stronger than diffusivity, i.e., $\beta>\alpha$. Therefore, a harmonic potential ($\beta=2$) is strong enough to control the enhanced diffusion for any $\alpha\leq2$. In this case, the system is ergodic in terms of the observable occupation fraction.
But for a more energetic particle, such as L\'{e}vy flight with divergent MSD, a deeper potential well is needed \cite{DybiecSokolovChechkin:2011}. If the potential well is not deep enough, the steady solution can not be normalized; for some discussions on the observables of interest, the time-dependent solution should be considered. In addition, infinite-ergodic theorem helps us to find the relation between time averages and ensemble averages, as well as the distribution of time averages.

For the case of $\beta=\alpha$, the Langevin equation with multiplicative noise is turned into a new Langevin equation with additive noise (i.e., the Brownian particle in a logarithmic potential) through the variable substitution. This way looks similar to the treatment of a Langevin equation with multiplicative noise $D(x)$ in non-Stratonovich sense (e.g., It\^{o} or H\"{a}nggi-Klimontovich sense) in a recent paper \cite{LeibovichBarkai:2018}.
Compared to it, we regularize $D(x)$ at the origin and investigate the effects of the large-$x$ behavior of $D(x)$ and $U(x)$ on the ergodic dynamics.
For different cases: $\beta=\alpha$ and $\beta<\alpha$, we use different kinds of scaling ansatz to obtain the scaling properties of the time-dependent solution $p(x,t)$ and even an analytic solution.

\section*{Acknowledgments}
This work was supported by the National Natural Science Foundation of China under grant no. 11671182, and the Fundamental Research Funds for the Central Universities under grant no. lzujbky-2018-ot03.

\appendix
\section{Relation between the space-dependent diffusivity and the distribution of waiting time}\label{APPa}
We first consider the coupled Langevin equation
\begin{equation}\label{Appa1}
    \frac{dx(s)}{ds}=\xi(s),  \quad  \frac{dt(s)}{ds}=\eta(s),
\end{equation}
where $\xi(t)$ and $\eta(t)$ are independent; it can describe a specific CTRW model \cite{Fogedby:1994}, and the subordinator $t(s)$ decides the distribution of the waiting times of CTRW model, of which the characteristic function is assumed to be
\begin{equation}\label{characteristic}
  \langle e^{-\lambda t(s)}\rangle = e^{-s\Phi(\lambda)}.
\end{equation}
By combining the two sub-equations in \eqref{Appa1}  \cite{CairoliBaule2:2015}, the subordinated process $y(t):=x(s(t))$ can be rewritten as a new Langevin equation \eqref{APPa2}. In fact,
\begin{equation}
  \begin{split}
    y(t)&=\int_0^{s(t)} \xi(\tau) d \tau   \\
    &= \int_0^{\infty} \delta(s-s(t))\int_0^s \xi(\tau)d\tau ds   \\
    &= \int_0^{\infty} -\frac{\partial}{\partial s}\Theta(t-t(s)) \int_0^s \xi(\tau)d\tau ds   \\
    &= \int_0^{\infty} \Theta(t-t(s)) \xi(s) ds,
  \end{split}
\end{equation}
where we technically add the $\delta$-function in the second line,  use a relation:  $ \Theta(s-s(t))=1-\Theta(t-t(s))$  between inverse subordinator $s(t)$ and subordinator $t(s)$ \cite{BauleFriedrich:2005} in the third line, and make integration by parts in the last line.
Therefore, $y(t)$ satisfies
\begin{equation}\label{APPa2}
  \frac{dy(t)}{dt}=\bar\xi(t), \quad \bar\xi(t)=\int_0^{\infty} \xi(s)\delta(t-t(s)) ds,
\end{equation}
where $\bar\xi(t)$ can be regarded as a new noise of the subordinated process $y(t)$; and letting $\xi(t)$ be white noise leads to the  correlation function of $\bar\xi(t)$ as
\begin{equation}
  \langle \bar\xi(t_1)\bar\xi(t_2)\rangle = \int_0^{\infty} \langle \delta(t_1-t(s))\delta(t_2-t(s)) \rangle ds,
\end{equation}
 the concrete form of which can be obtained from its Laplace transform ($t_1\rightarrow\lambda_1,t_2\rightarrow\lambda_2$) by using the characteristic function of $t(s)$ in \eqref{characteristic}, i.e.,
\begin{equation}
  \langle \bar\xi(\lambda_1)\bar\xi(\lambda_2)\rangle = \int_0^{\infty}  e^{-s\Phi(\lambda_1+\lambda_2)} ds  = \frac{1}{\Phi(\lambda_1+\lambda_2)}.
\end{equation}
Taking the inverse Laplace transform in the above equation, when $\Phi(\lambda)=\lambda/(2D(x))$, we obtain the same form of the correlation function of multiplicative white noise
\begin{equation}\label{newnoise}
\begin{split}
    \langle \bar\xi(t_1)\bar\xi(t_2)\rangle &= \mathcal{L}^{-1}[\Phi(\lambda)^{-1}]\, \delta(t_1-t_2)  \\
    &=2D(x) \delta(t_1-t_2).
\end{split}
\end{equation}
This specific $\Phi(\lambda)$ in \eqref{characteristic} signifies that the waiting time in the associated CTRW model obeys exponential distribution with rate $2D(x)$, i.e.,
\begin{equation}
  \phi(\tau)=2D(x)e^{-2D(x)\tau}.
\end{equation}
The space-dependent rate $2D(x)$ denotes the number of renewals per unit time, which means that for large $D(x)$ the mean of waiting time is small and thus the diffusion is enhanced and vice versa.
Note that no external fields are considered here for simplicity, since we aim to derive the new noise $\bar\xi(t)$ in \eqref{newnoise} with the space-dependent correlation. Actually, the Langevin equation \eqref{model2} is equivalent to the first equation in \eqref{APPa2} with an external field  $f(y)$.

\section{Scaling ansatz for $\beta=\alpha$}\label{APPb}
We employ the scaling ansatz with the scaling function $g(z)$ for $p(y,t)$ of \eqref{Eq_y} as:
\begin{equation}\label{Appb1}
p(y,t)=t^{-\gamma}g(z), \quad z=y/t^\gamma.
\end{equation}
The Fokker-Planck equation corresponding to the Langevin equation of $y(t)$ \eqref{Eq_y} is
\begin{equation}\label{Appb2}
  \frac{\partial p(y,t)}{\partial t} = \frac{\partial}{\partial y}\frac{A}{y}p(y,t) + \frac{1}{2}\frac{\partial^2}{\partial y^2}p(y,t).
\end{equation}
Substituting \eqref{Appb1} into the Fokker-Planck equation \eqref{Appb2}, we get the equation of the scaling function $g(z)$ after replacing the variable $y$ with $z$:
\begin{equation}\label{Appb3}
\begin{split}
    &t^{-1-\gamma}(-\gamma g(z)-\gamma zg'(z)) \\
    &~~~=t^{-3\gamma} \left[\frac{1}{2}g''(z)+\frac{A}{z}g'(z)-\frac{A}{z^2}g(z) \right].
\end{split}
\end{equation}
Comparing the terms with $t$, we find the scaling solution can be obtained by eliminating the terms with $t$ when $\gamma=1/2$.
Then for small $z$ the left-hand side of Eq. \eqref{Appb3} tends to zero compared to the right-hand side, and it becomes
\begin{equation}\label{Appb4}
  \frac{1}{2}g''(z)+\frac{A}{z}g'(z)-\frac{A}{z^2}g(z)=0,
\end{equation}
and the solution of \eqref{Appb4} is
\begin{equation}
  g(z)\propto |z|^{-2A}  ~~~\textrm{and} ~~~    g(z)\propto |z|.
\end{equation}
The latter one is rejected since a PDF must decay at the infinity. To obtain the concrete coefficient in front of $|z|^{-2A}$, the complete form of $g(z)$ including the scaling for large $z$ should be derived firstly.
For this, we assume $g(z)=N_g|z|^{-2A}h(z)$, and obtain
\begin{equation*}
  h''(z)+(z-2A/z)h'(z)+(1-2A)h(z)=0,
\end{equation*}
the solution of which is $\exp(-z^2/2)$. Then the complete form of $g(z)$ is
\begin{equation}\label{gz}
  g(z)=N_g |z|^{-2A} \exp(-z^2/2)
\end{equation}
with the normalized parameter $N_g$ being
\begin{equation*}
  N_g=2^{A-1/2}[\Gamma(1/2-A)]^{-1}.
\end{equation*}
Substituting \eqref{gz} into \eqref{Appb1} yields
\begin{equation}\label{Appb5}
\begin{split}
    p(y,t)&= N_g t^{A-\frac{1}{2}} |y|^{-2A} \exp(-y^2/(2t))  \\
    &\simeq N_g t^{A-\frac{1}{2}} |y|^{-2A},
\end{split}
\end{equation}
when $|y|\ll t^{1/2}$.


\section{Scaling ansatz for $\beta<\alpha$}\label{APPc}
Considering the exact solution in \eqref{pdf1} for a free particle in a heterogeneous media,
we use the scaling ansatz as
\begin{equation}\label{Appc1}
\begin{split}
    p(x,t)&=\frac{1}{\sqrt{4\pi D_0t}} \exp\left(-\frac{|x|^{2-\alpha}}{(2-\alpha)^2D_0t}\right) \cdot p_0(x)  \\[3pt]
    &=: q(x,t) \cdot p_0(x)
\end{split}
\end{equation}
to keep the same shape of the solution in \eqref{pdf1} for long times. Note that we remain the terms depending on time $t$ but hide the term $|x|^{-\alpha/2}$ from \eqref{pdf1} by putting it into the auxiliary function $p_0(x)$.
Substituting the $p(x,t)$ \eqref{Appc1} into the Fokker-Planck equation \eqref{FFP}, we obtain
\begin{equation}
  \begin{split}
    &p_0(x) \frac{\partial q}{\partial t}  = -q \frac{\partial}{\partial x}[f(x)p_0(x)]-f(x)p_0(x)\cdot \frac{\partial q}{\partial x}  \\
    &+ \frac{\partial}{\partial x} \left[D(x)p_0(x) \frac{\partial q}{\partial x}\right]
    +q\frac{\partial}{\partial x} \left[\sqrt{D(x)}\frac{\partial}{\partial x}\sqrt{D(x)}p_0(x)\right] \\
    &+\sqrt{D(x)}\frac{\partial}{\partial x}\left[\sqrt{D(x)}p_0(x)\right]\cdot \frac{\partial q}{\partial x} .
  \end{split}
\end{equation}
For large $t$, the terms containing $\partial q/\partial t$ and $\partial q/\partial x$ can be omitted, compared with $q$. The leading terms remained form the equation
\begin{equation}
  -q \frac{\partial}{\partial x}[f(x)p_0(x)]+q\frac{\partial}{\partial x} \left[\sqrt{D(x)}\frac{\partial}{\partial x}\sqrt{D(x)}p_0(x)\right]=0,
\end{equation}
the solution of which is exactly the steady solution in \eqref{pst2}, i.e.,
\begin{equation}
  p_0(x)=p^{st}(x).
\end{equation}
So we finally get the time-dependent solution
\begin{equation}\label{Appc2}
    \lim_{t\rightarrow\infty}p(x,t)=\frac{1}{\sqrt{4\pi D_0t}} \exp\left(-\frac{|x|^{2-\alpha}}{(2-\alpha)^2D_0t}\right) \cdot p^{st}(x).
\end{equation}
Since for large $x$, Eq. \eqref{pst2} shows that
\begin{equation}
  p^{st}(x) \simeq \frac{N}{\sqrt{D_0}}|x|^{-\alpha/2},
\end{equation}
the coefficient $N$ can be determined through the normalization of $\lim_{t\rightarrow\infty}p(x,t)$ in \eqref{Appc2}, and
\begin{equation}
  N=\sqrt{D_0}.
\end{equation}


\bibliographystyle{aipnum4-1}
\bibliography{Reference}

\end{document}